# A Gamma/Electron Spectrometer with Logarithmic ADC based on LC Tank Ring-Down Oscillation Counting

---

**S. Abovyan, A. Chilingarian, D. Pokhsraryan, B. Sargsyan, T. Karapetyan, V.Danielyan**

*A. I. Alikhanyan National Lab (Yerevan Physics Institute),*
*Alikhanyan Brothers 2, Yerevan, Armenia AM0036*
*E-mail*: s.abovyan@aanl.am

ABSTRACT: This paper describes the operating principle, circuit implementation, and mathematical basis of a logarithmic analog-to-digital converter (ADC) that exploits the natural exponential decay of a damped LC tank oscillator. An input charge pulse excites a tuned LC resonator into free oscillation. A multi-stage high-speed amplifier chain buffers the decaying waveform, and a fast comparator counts the number of oscillation cycles that exceed a programmable threshold. Because the tank envelope decays exponentially, the cycle count is proportional to the natural logarithm of the input charge, yielding logarithmic compression over a wide dynamic range without explicit logarithmic circuitry. The upgraded unit incorporates a modern MCU with USB, Ethernet interfaces and SD card for collected data storage. The architecture is well suited to applications requiring large dynamic range in a compact, low-power implementation.

KEYWORDS: Logarithmic analog-to-digital converters; LC resonant circuits; Cosmic-ray detectors; Gamma-ray and electron spectrometers; Thunderstorm ground enhancements; Front-end electronics for particle detectors.

## 1. Introduction

Measuring signals that span several decades of amplitude with high resolution is a perennial challenge in instrumentation, nuclear electronics, and sensor readout systems. Conventional linear ADCs allocate equal quantization steps across the full input range, which is inefficient when signals of interest follow a logarithmic or exponential distribution. Dedicated logarithmic amplifiers can compress dynamic range before digitization, but they introduce nonlinearity errors and temperature drift that must be carefully managed.

An alternative, more elegant approach exploits the inherent exponential behavior of a damped resonator. When a charge pulse is deposited into a parallel LC tank, the tank oscillates freely at its resonant frequency, and the oscillation amplitude decays exponentially due to resistive losses. A fixed-threshold comparator then counts the number of oscillation cycles that remain above the threshold. Because the number of such cycles is proportional to the logarithm of the initial amplitude — which is itself proportional to the input charge — the cycle count forms a natural logarithmic digital representation of the input charge with no explicit log function required.

This technique is analogous in spirit to the Wilkinson ramp ADC used in nuclear instrumentation but replaces the linear ramp with an exponentially decaying oscillation, trading linear resolution for logarithmic compression.

## 2. Operating Principle

### 2.1 LC Tank Excitation

Consider a parallel LC tank with inductance L, capacitance C, and equivalent parallel resistance R representing core and winding losses. When a charge is deposited across the tank, the initial voltage across the capacitor is:

$$V_0 = Q / C$$

The tank then oscillates at its natural resonant frequency:

$$f_0 = 1 / (2\pi \sqrt{(LC)})$$

For the component values in the described circuit (L = 33 μH, C = 330 pF):

$$f_0 = 1 / (2\pi \sqrt{(33\times10^{-6} \times 330\times10^{-12})}) \approx 1.52 \text{ MHz}$$

In this specific case the frequency is limited to the 1.5 - 2 MHz range, because if the duration of the input pulses (for PMT pulses this is 300–500 ns) exceeds a quarter of the resonance frequency period, the proportionality of the number of output pulses to the logarithm of the input signal is disrupted.

### 2.2 Exponential Envelope Decay

The oscillation is damped by resistive losses, and the envelope of the tank voltage follows:

$$V(t) = V_0 \cdot exp(-t/\tau) \cdot cos(2\pi f_0 t)$$

where the decay time constant is $\tau$ = 2RC = 2L / $R_{series}$. The quality factor of the tank determines how many oscillation cycles occur before the amplitude becomes negligible:

$$Q_{ta(k} = R / (\omega_0 L) = \pi f_0 \tau$$

### 2.3 Logarithmic Cycle Counting

A comparator with a fixed threshold voltage $V_{thr}$ examines the oscillating waveform and produces a digital pulse for each cycle that crosses the threshold. The envelope reaches $V_{thr}$ at time:

$$t_{stop} = \tau \cdot ln (V_0 / V_{thr})$$

The number of oscillation cycles counted to this point is:

$$N = t_{stop} / T_0 = (\tau / T_0) \cdot ln (V_0 / V_{thr})$$

Substituting $V_0$ = Q/C:

$$N = (\tau / T_0) \cdot ln (Q / (C \cdot V_{thr}))$$

Since $\tau$, $T_0$, C, and $V_{thr}$ are all fixed circuit parameters, the cycle count N is strictly proportional to the natural logarithm of the input charge Q. This is the fundamental result: the circuit is a logarithmic charge-to-digital converter with no need for explicit log-function computation.

## 3. Channel Circuit Description

LADC unit consists of 4 identical analog signal channels. The implemented analog part of circuit of a single channel consists of four functional blocks arranged in cascade: a V-to-I converter, two intermediate gain/limiting stages, and a final comparator.

### 3.1 Stage 1 — Voltage-to-Current converter (U21, Q5)

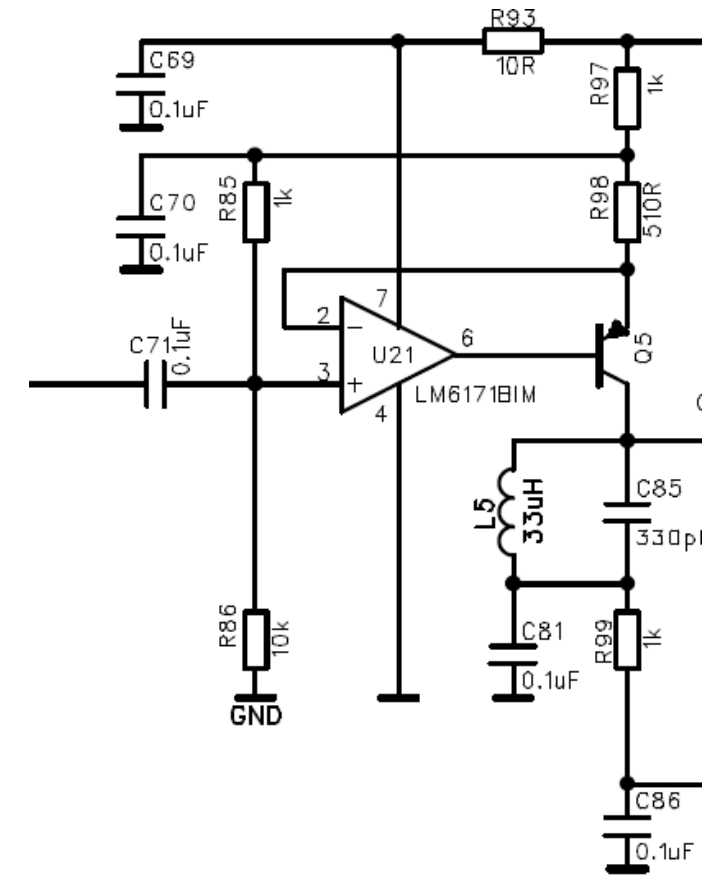

The input pulse arrives at connector and is AC-coupled through C71 (0.1 µF) into the non-inverting input of U21, a LM6171BIM high-speed operational amplifier. Resistor R86 (10 kΩ) provides a DC return path to ground. The LM6171BIM is selected for its 100 MHz gain-bandwidth product and low distortion, ensuring the 1.52 MHz oscillation is amplified with minimal phase error.

Transistor Q5 (NPN) is connected as an output stage driven by the op-amp output through R93 (10 Ω), providing additional current drive capability. Resistors R97 (1 kΩ) and R98 (510 Ω) form a feedback voltage divider from Q5's collector back to the inverting input of U21, stabilizing closed-loop gain.

The transconductance is $G_m = \frac{1}{R_{98}} = \frac{1}{510\,\Omega} \approx 2$ mA/V.

This value is selected so that for the standard shape of PMT pulse, the peak voltage of the first half-wave on the oscillating LC tank would be equal to the amplitude of the input pulse. Since the maximum amplitude of the input pulse is set in the preamplifier to the level of approximately 7V, the same amplitude is obtained at the entrance of Operational Amplifier (OA) U21. The greater maximal voltage values cause the limitation of oscillations amplitudes due to the sharp increase of the U2 input current, whereas with the smaller values the dynamic range of converter decreases due to an increase in the portion of OA inherent noise, the Radio-Frequency Interferences (RFI) from the radio stations and other kinds of Electromagnetic Interferences (EMI).

The LC tank formed by L5 (33 µH) and C85 (330 pF) defines the oscillation frequency and the decay time constant τ that sets the logarithmic scale factor.

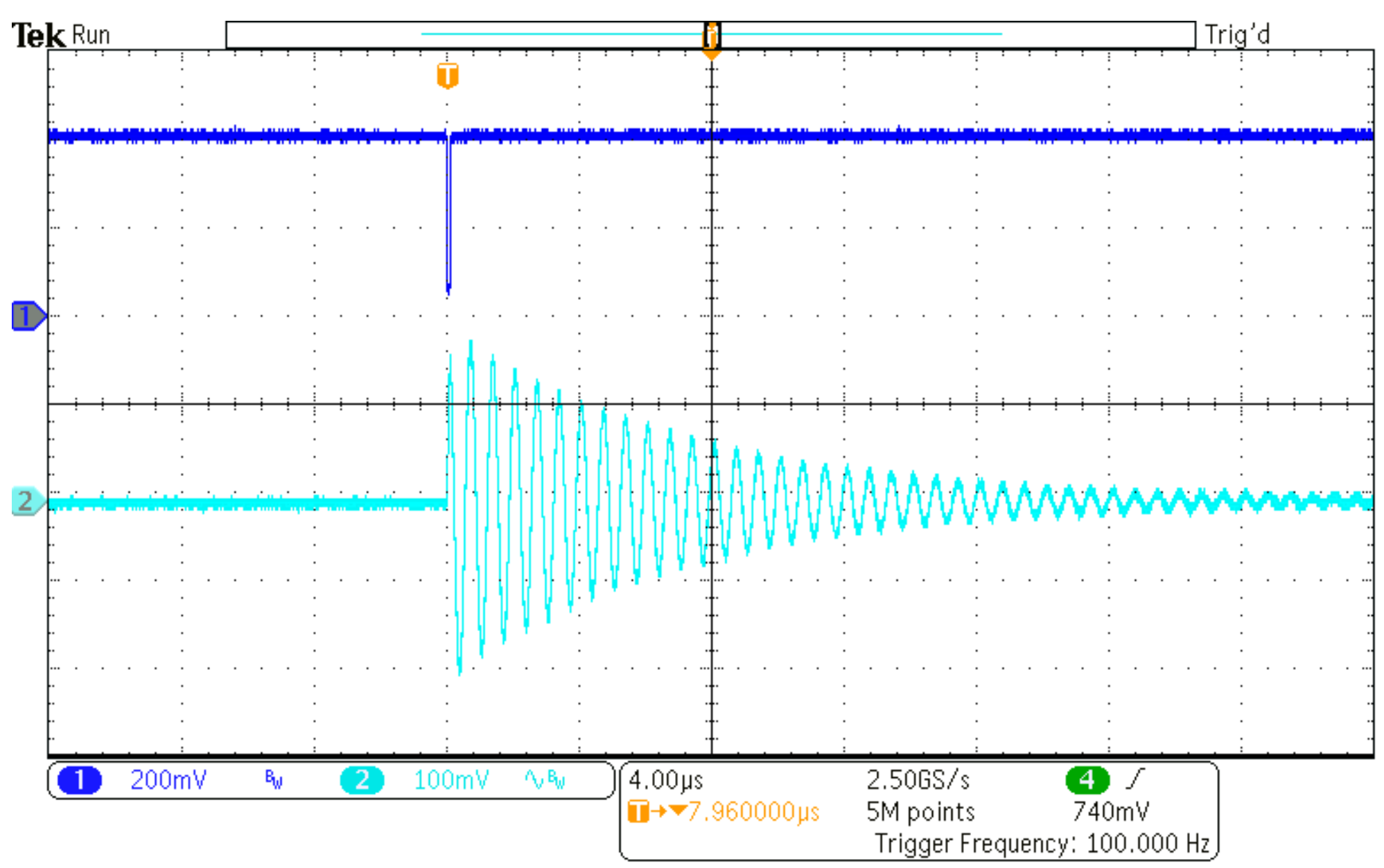

***An input signal (dark blue) and damping oscillations (light blue)***

The oscillation amplitude on the RLC tank is proportional to the area of the pulse shape, if its duration is shorter than one fourth of the tank resonance period, i.e. $V_O = \frac{1}{C}\int I_{pls} d$

### 3.2 Stage 2 — Intermediate Amplifier with Clipping (U29)

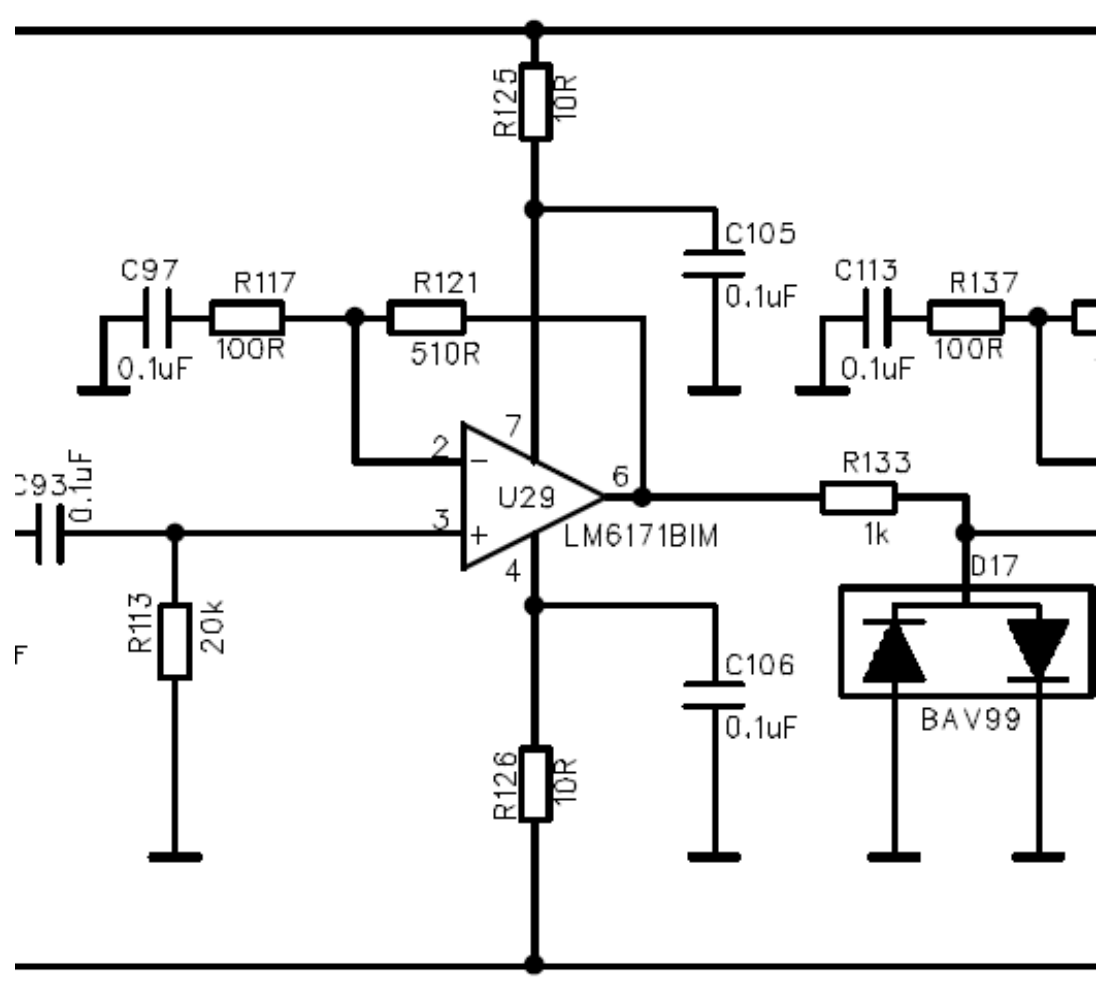

The oscillating signal is AC-coupled through C93 (0.1 μF) into a second LM6171BIM stage (U29), with feedback set by R117 (100 Ω) and R121 (510 Ω). Diode D17 (BAV99 dual switching diode) acts as an amplitude limiter across the feedback path. During the first oscillation cycles — when the tank amplitude is large — D17 clamps the output, preventing hard saturation of U29 and ensuring rapid recovery to faithfully track later, smaller-amplitude cycles. Without this limiting, amplifier recovery time would suppress early cycles and corrupt the count.

### 3.3 Stage 3 — Intermediate Amplifier with Schottky Limiting (U33)

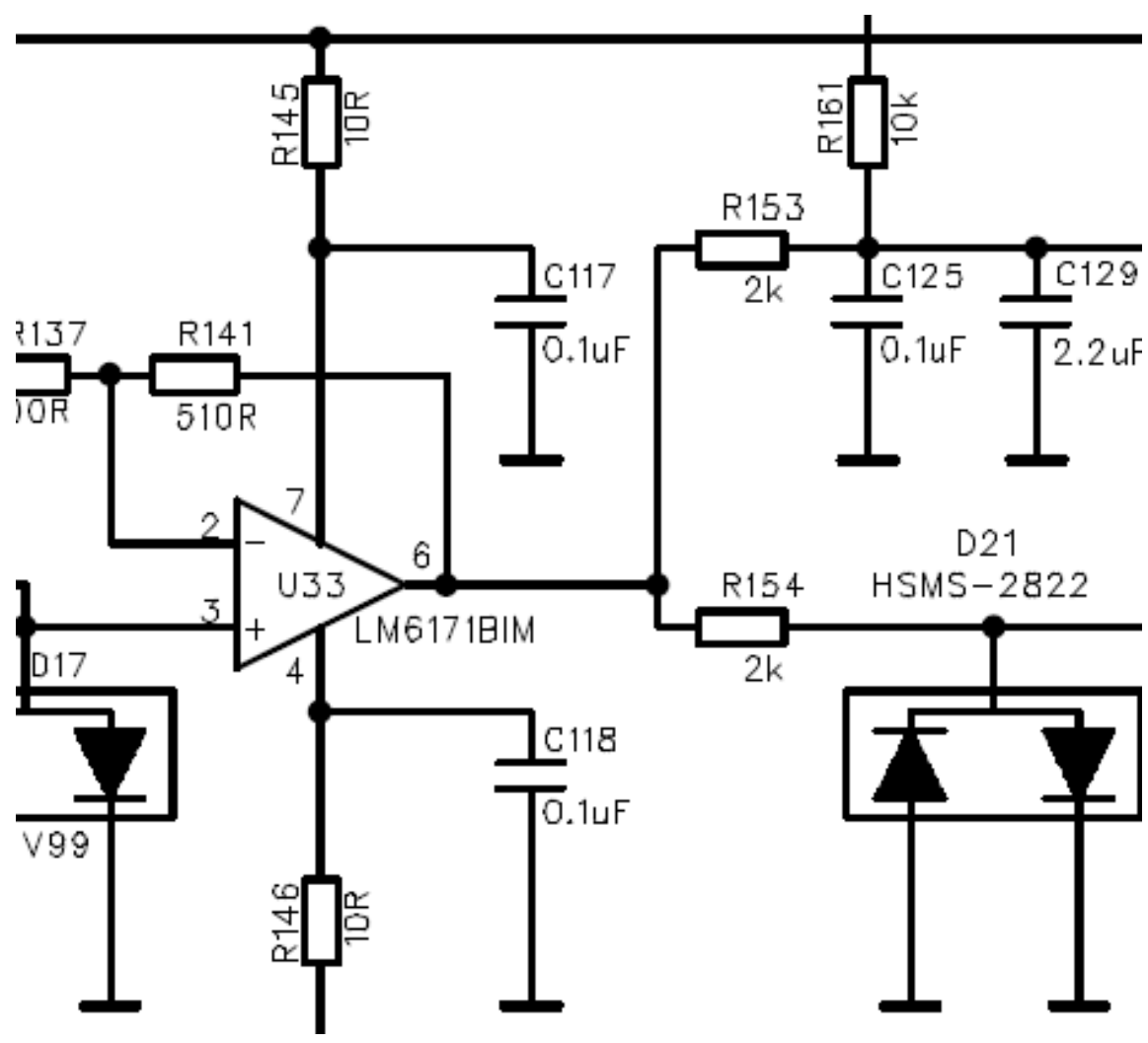

A third LM6171BIM (U33) provides additional gain with R137/R141 feedback. Diode D21 (HSMS-2822 Schottky pair) performs the same limiting function as D17 but with a lower forward voltage drop and faster switching, optimized for RF operation at 1.52 MHz. Together, D17 and D21 ensure all three amplifier stages operate in a controlled gain regime throughout the ring-down, from large early cycles to small late cycles near the threshold.

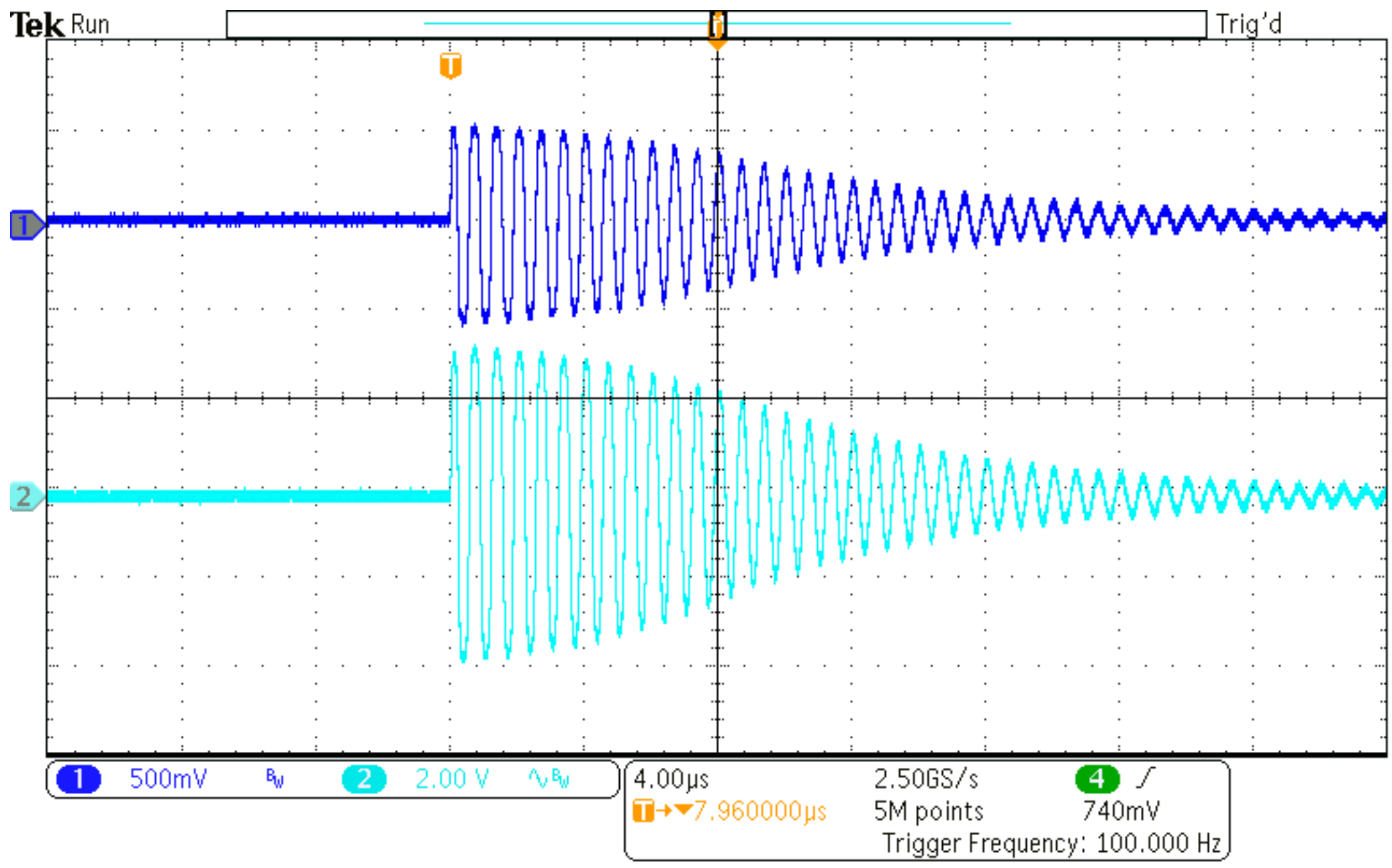

***The output signal of the first stage of amplifier (dark blue) and input signal of comparator (light blue)***

### 3.4 Stage 4 — Comparator (U37)

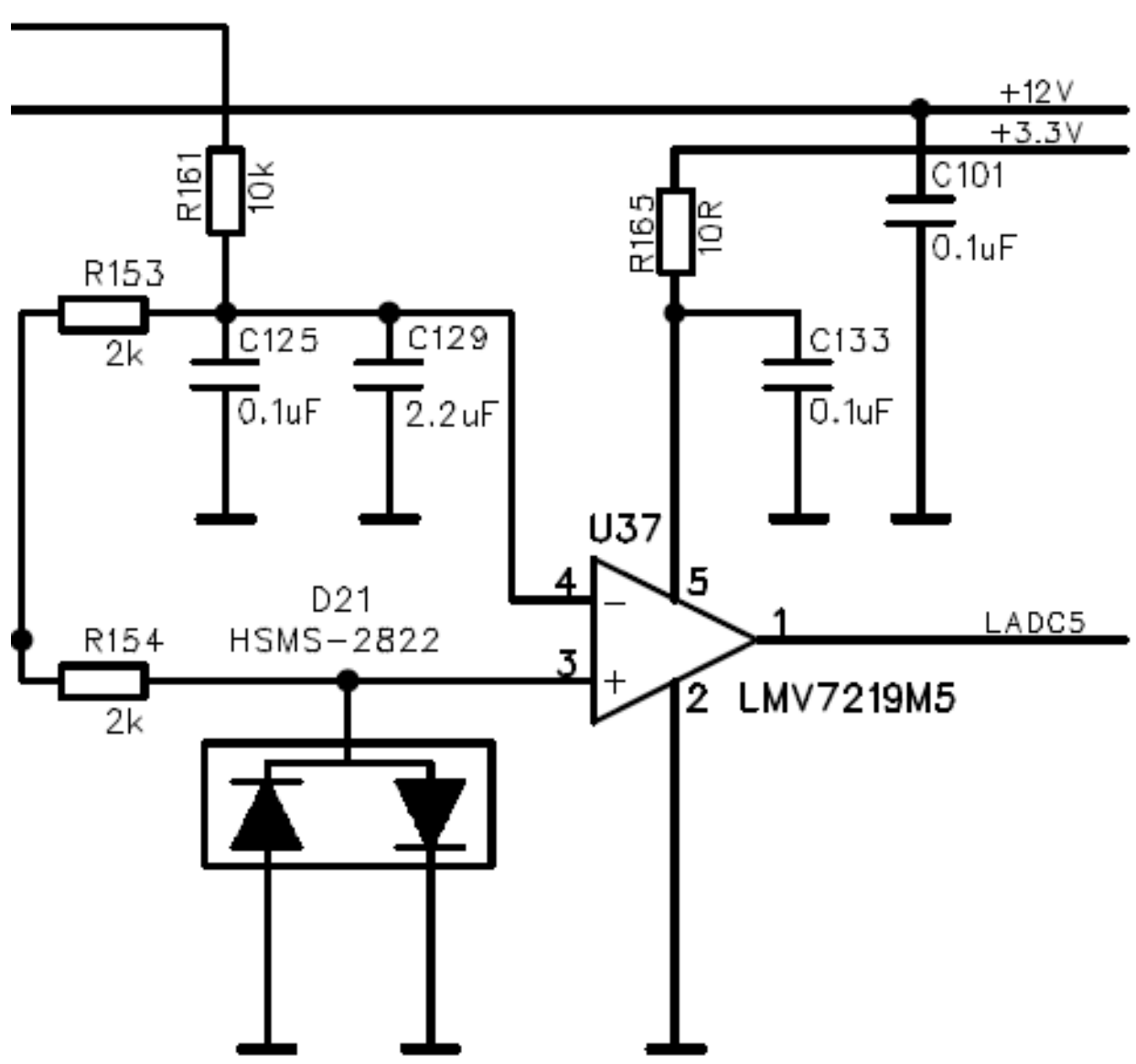

The final stage uses a LMV7219M5 high-speed comparator with 7 ns propagation delay. The non-inverting input receives the amplified oscillating signal; the inverting input is biased by R153/R154 (both 2 kΩ) connected to the THR5 threshold line. Adjusting THR5 sets $V_{thr}$, controlling the low end of the dynamic range. The comparator output (LADC5) is a digital pulse train whose pulse count encodes log(Q).

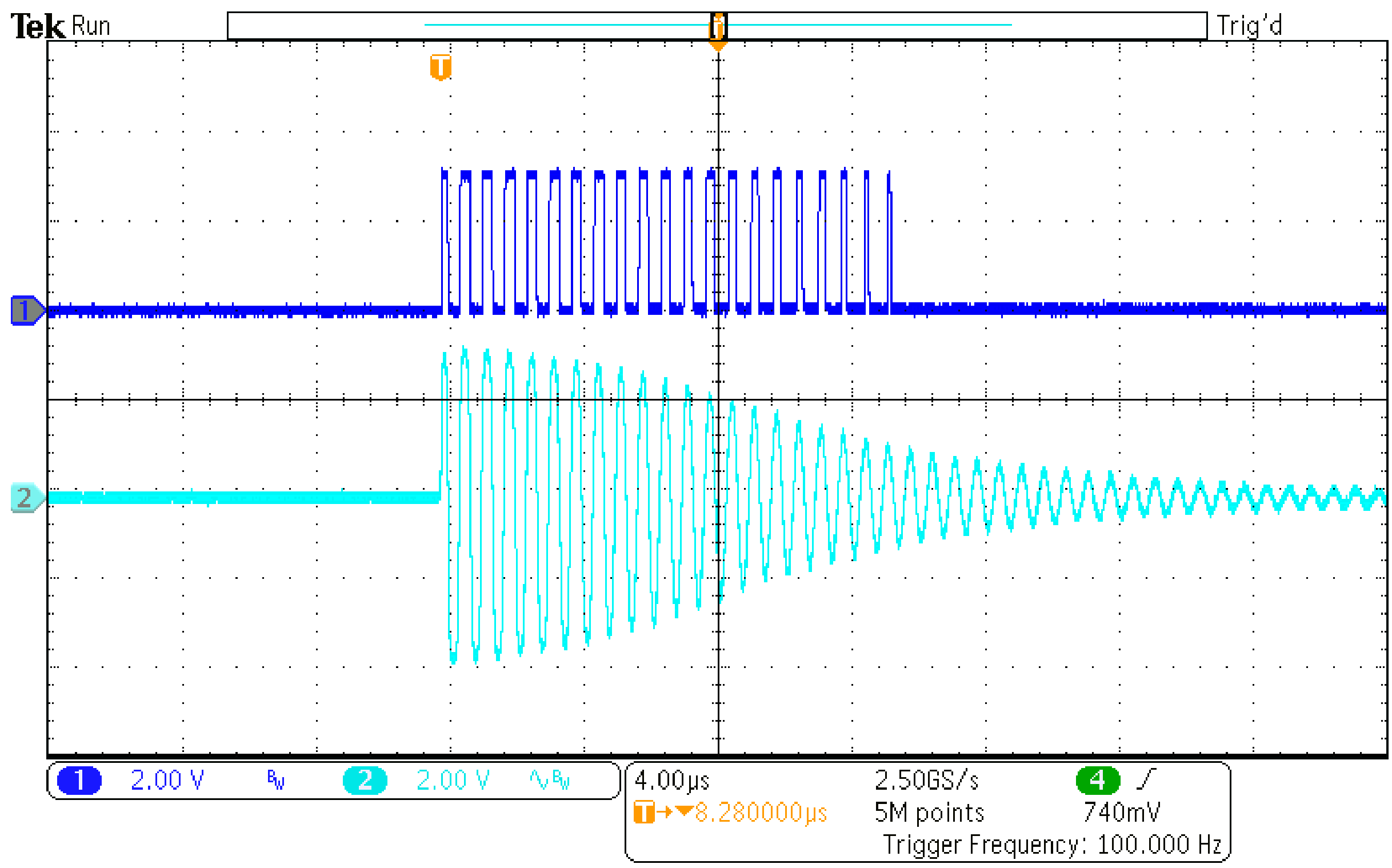

**Figure 1 Input (light blue) and output (dark blue) of comparator**

## 4. Component Selection Rationale

| Component | Value | Role |
|---|---|---|
| **L5** | 33 µH | Sets resonant frequency and decay time constant |
| **C85** | 330 pF | Sets resonant frequency |
| **U21, U29, U33** | LM6171BIM | 100 MHz GBW high-speed amplification |
| **Q5** | NPN BJT | Current buffer, increases output drive |
| **D17** | BAV99 | Dual switching diode limiter |
| **D21** | HSMS-2822 | Schottky RF limiter, low forward voltage |
| **U37** | LMV7219M5 | 7 ns comparator, digital output |
| **R86** | 10 kΩ | DC bias for op-amp non-inverting input |
| **R93, R125, etc.** | 10 Ω | Series isolation, prevent parasitic oscillation |
| **THR5** | Adjustable | Sets detection threshold $V_{thr}$ |

## 5. Logarithmic Scale Factor and Calibration

The scale factor relating cycle count to log(Q) is identically the tank quality factor:

$$S = dN / d\,(\ln Q) = \tau / T_0 = f_0 \cdot \tau = Q_t$$

A high-Q tank gives more counts per decade of input charge, improving resolution. A low-Q tank compresses the count range. Deliberate damping resistors in parallel with the tank allow $\tau$ to be tuned independently of $f_0$, adjusting the logarithmic scale factor without changing the oscillation frequency.

Calibration requires only two known input charges $Q_1$ and $Q_2$, yielding counts $N_1$ and $N_2$:

$$\tau = (N_2 - N_1) \cdot T_0 / \ln (Q_2 / Q_1)$$

$$C \cdot V_{thr} = Q_1 \cdot \exp (-N_1 T_0 / \tau)$$

These two parameters fully characterize the converter.

## 6. Noise and Threshold Considerations

The minimum detectable charge is set by the condition that the initial tank voltage $V_0$ exceeds the referred-input noise of the amplifier chain plus the threshold:

$$Q_{min} = C \cdot (V_{thr} + V_{noise,\ rms})$$

The LM6171BIM has a voltage noise density of approximately 14 nV/√Hz. Referred to the 1.52 MHz bandwidth of the tank and across three amplifier stages, the noise floor remains comfortably below a practically chosen $V_{thr}$. The 7 ns propagation delay of the LMV7219 avoids timing jitter that would smear the cycle count at high oscillation frequencies.

At the high end, $Q_{max}$ is limited by amplifier recovery from the large initial clamped cycles. The BAV99 and HSMS-2822 limiters prevent U29 and U33 from being driven deeply into saturation, keeping recovery within one or two oscillation periods, and avoiding missed counts at the start of the ring-down.

## 7. Dynamic Range

### 7.1 Key Parameters

- L5 = 33 μH, C85 = 330 pF → $f_0$ = 1.52 MHz, $T_0$ = 658 ns
- Stage 1: V-to-I converter, Gm = 1/R98 = 1/510Ω ≈ 2 mA/V; by design $V_{0t} = V_{in}$ (preamplifier maximum ≈7 V)
- Stages 2 and 3: voltage gain ×6.1 (U29) ×6.1 (U33) ≈ ×37.2 combined
- BAV99 clamp voltage ≈ ±0.7 V; HSMS-2822 ≈ ±0.25 V (limiters, not detectors)
- LMV7219 minimum usable threshold ≈ 4 mV (TLV5626 8-bit DAC, $V_{ref}$ ≈1 V → step ≈4 mV)

**Tank Q is the dominant uncertainty.** The actual Q of the physical inductor directly sets the scale factor. A ±20% variation in Q shifts every count reading by ±20%, which is why in-situ calibration via the TLV5626 threshold sweep is essential.

**THR trades sensitivity for noise immunity.** Raising $V_{thr}$ reduces $Q_{min}$ sensitivity but suppresses noise-triggered false counts, effectively narrowing the dynamic range from the bottom end.

## 7.2 Measurements of Dynamic Range

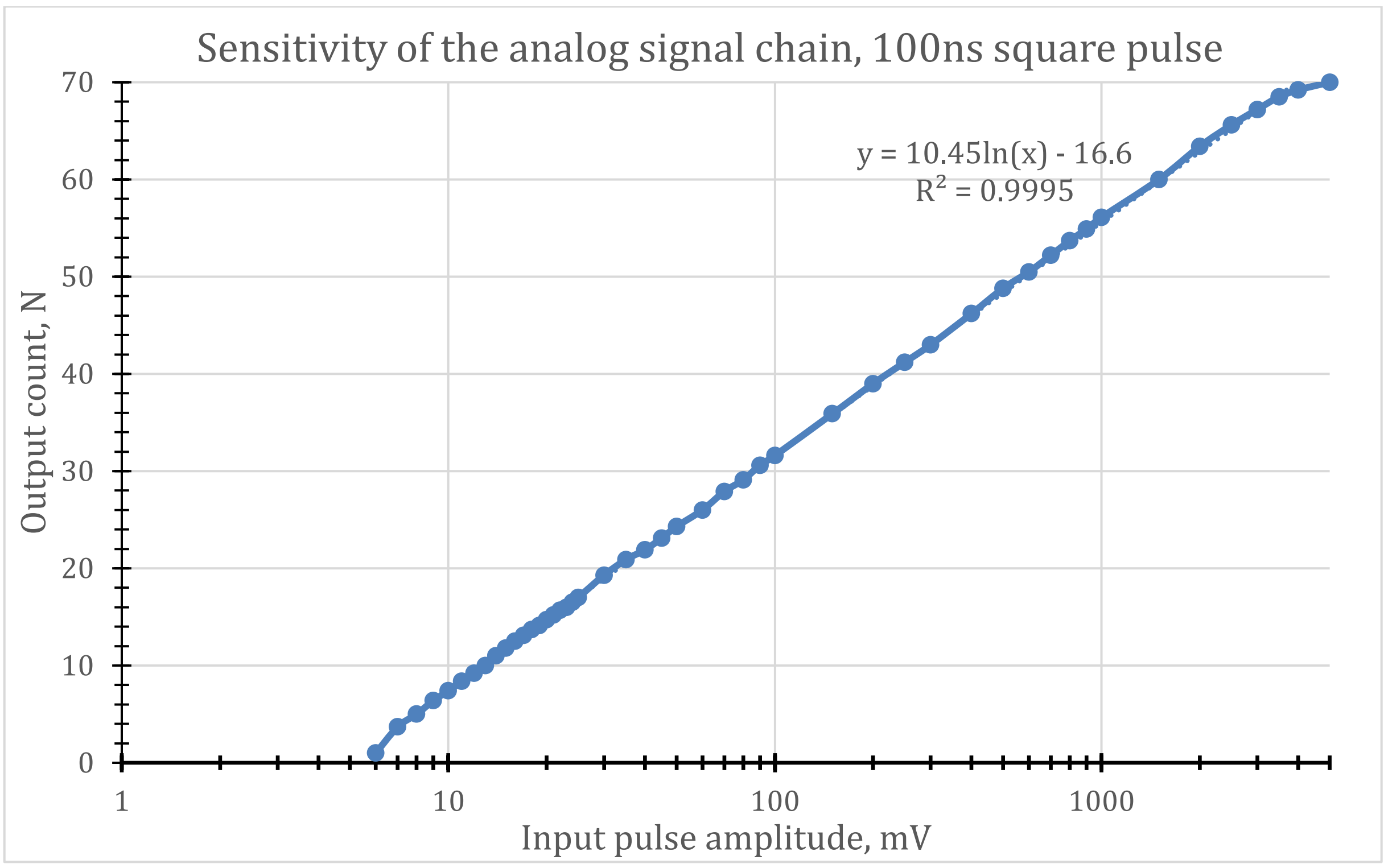

The plots show the measured transfer function of the LADC analog chain driven by a 100 ns square pulse, presented on logarithmic amplitude axes. The logarithmic axis plot is the key diagnostic: if the circuit behaves as a true logarithmic converter, the data should fall on a straight line when plotted against log(V_in), and it does so with exceptional fidelity across nearly three decades of input amplitude. The fitted equation $N = 10.45 \cdot \ln(V) - 16.6$ achieves $R^2 = 0.9995$, confirming that the LC ring-down counting mechanism produces a rigorously logarithmic response rather than an approximation.

## 7.3 Extracting Circuit Parameters from the Fit

The two constants in the fit equation directly yield physical circuit parameters. The slope $S = 10.45$ is the logarithmic scale factor $\tau/T_0$, from which the effective tank quality factor follows immediately: $Q_{tank} = \pi \times S = \pi \times 10.45 \approx$ **33**. This is the measured Q of the actual 33 µH inductor at 1.52 MHz on the PCB — lower than the ideal assumed value of 50, which is typical for a wound inductor with copper-loss and board parasitic contributions. The corresponding ring-down time constant is $\tau = T_0 \times S = 658$ ns $\times$ 10.45 ≈ **6.9 µs**, meaning the tank amplitude falls to 1/e in roughly ten oscillation periods.

The intercept −16.6 encodes the effective comparator threshold referred back to the input. Setting $N = 0$ gives the minimum detectable voltage:

$$V_{min} = e^{16.6/10.45} \approx \mathbf{5}\ \text{mV}$$

This is consistent with the TLV5626 DAC threshold setting and the LMV7219 comparator offset, validating that the measured minimum corresponds to the design intent rather than noise-floor saturation.

The data in image 1 shows a saturation starting around 3500 mV. Extrapolating the fit to the 7 V design maximum input gives N_theoretical ≈ 76, suggesting the firmware counter at 80 to capture the full range intended for PMT-shaped pulses.

The measured usable range for the 100 ns test pulse runs from approximately **5 mV to 3500 mV** (where the counter saturates), giving a measured dynamic range of:

$$DR_{measured} = 20\log_{10}\left(\frac{3500}{5}\right) \approx \mathbf{57}\ \text{dB}$$

## 7.4 Amplitude-Dependent Dead Time

In a conventional ADC (flash, SAR, pipeline), dead time is fixed and independent of the input. In a Wilkinson ADC it is proportional to the input amplitude. In this LC ring-down system it is proportional to the **logarithm** of the input amplitude — because the conversion time *is* the ring-down duration, which *is* the quantity being measured:

$$T_{dead}(V_0) = N \times T_0 = \tau \cdot \ln\left(\frac{V_0}{V_{thr}}\right)$$

With the measured parameters ($\tau = 6.9$ µs, $T_0 = 658$ ns):

| Input amplitude | N | T_dead |
|---|---|---|
| 5 mV (minimum) | 1 | **0.66 µs** |
| 50 mV | ~20 | **13 µs** |
| 500 mV | ~38 | **25 µs** |
| 3500mV (saturation) | ~70 | **46 µs** |

The dead time spans nearly two orders of magnitude across the dynamic range — a property unique to this architecture.

### 7.4.1 Formal Definition

The total dead time per event has three additive components:

$$T_{dead} = \underbrace{T_{pulse}}_{\text{input duration}} + \underbrace{N \cdot T_0}_{\text{ring-down (conversion)}} + \underbrace{T_{reset}}_{\text{digital processing}}$$

- **T_pulse** (~100–500 ns): the duration of the input pulse itself, during which the tank is being driven and cannot be cleanly re-excited. For PMT pulses this is 300–500 ns; for the 100 ns test pulses it is negligible.

- **N · $T_0$**: the ring-down duration — the dominant and amplitude-dependent term.
- **T_reset**: Teensy latching the count, resetting the comparator state, and asserting readiness. With current implementation it is 1us.

Since N is the direct observable output, the dead time correction can be computed on a **per-event basis** without any additional measurement:

$$T_{dead}(event) = N_{event} \times 658\,\text{ns} + T_{reset}$$

This is a rare and convenient property — the system self-reports its own dead time.

### 7.4.2 The System is Paralyzable

There is no input gating or blanking during the ring-down — the input at remains continuously connected through C71 and R86. If a second pulse arrives at time t after the first, while the tank is still ringing with amplitude:

$$V_{residual}(t) = V_0 \cdot e^{-t/\tau}$$

the new excitation adds vectorially to the residual oscillation. Depending on the relative phase:

- **In-phase arrival**: amplitudes add → longer ring-down than either pulse alone
- **Out-of-phase arrival**: partial cancellation → shorter ring-down, corrupted count

This is the definition of a **paralyzable (Type II / extendable) dead-time model**. At high event rates, pile-up does not simply cause event loss — it actively corrupts the amplitude measurement of both pulses, which is more damaging than simple dead-time loss.

### 7.4.3 Practical Dead-Time Correction

For a paralyzable system with amplitude-dependent dead time, the standard non-paralyzable correction $n = m/(1 + m\cdot\tau_d)$ does not apply directly. A more appropriate approach uses the fact that T_dead scales with N:

$$m = \frac{n}{1 - n \cdot \langle N \rangle \cdot T_0}$$

where ⟨N⟩ is the mean count per event for the current source conditions. For a known spectrum, this can be computed from the accumulated histogram. Alternatively, since T_dead is available event-by-event, a **time-stamped event-by-event correction** using the Teensy's NTP-disciplined real-time clock is possible without any approximation.

### 7.4.4 Implication for Cosmic Ray Detection

For a typical cosmic ray muon rate at sea level (~100–200 $m^{-2}s^{-1}sr^{-1}$), and a detector area of ~0.1 $m^2$, the single-channel event rate is of order 1–5 Hz. At N = 70 (maximum pulse), T_dead ≈ 46 µs. The fractional dead time is:

$$\delta = r \cdot T_{dead} \approx 5\,\text{Hz} \times 46\,\mu\text{s} = 0.023\%$$

This is entirely negligible. The amplitude-dependent dead time only becomes a design concern if the same hardware is repurposed for higher-rate applications — accelerator experiments, high-activity source calibration — where rates could reach kHz or above.

# 8. Advantages and Limitations

## 8.1 Advantages

- Logarithmic compression is obtained without log amplifiers, diode log converters, or look-up tables — it emerges directly from the exponential physics of LC decay.
- The oscillation frequency provides an intrinsic conversion clock; no separate ADC clock is required.
- The core conversion element is a passive LC tank. The remaining circuitry is standard op-amp and comparator design.
- The threshold THR and tank damping together provide two independent adjustment handles for sensitivity and scale factor.

## 8.2 Limitations

- Conversion time is proportional to τ multiplied by the number of counts. High-Q tanks improve resolution but increase conversion time, limiting throughput.
- Tank Q varies with temperature due to inductor resistance, causing drift in the scale factor. Temperature compensation or periodic calibration is required in precision applications.
- Any re-excitation of the tank after the initial pulse — from supply transients or crosstalk between channels — will corrupt the count. Careful layout and supply decoupling are essential.
- A lookup table or software linearisation step is required if a linear output representation is needed downstream.

# 9. System Context: The LADC 4-Channel Cosmic Ray Module

The full module is a 4-channel logarithmic analog-to-digital converter intended for cosmic ray particle detector readout, where signal amplitudes routinely span three to four decades of dynamic range.

Each of the four channels (PIN5–PIN8) is electrically identical, with its own V-to-I converter, LC tank, two-stage LM6171BIM amplifier chain, BAV99/HSMS-2822 limiting diodes, and LMV7219 comparator. The four digital outputs (LADC5–LADC8) feed a shared digital back end described in the following subsections. The modular, replicated structure minimizes interchannel crosstalk and allows each channel to be independently calibrated via a per-channel programmable threshold.

### 9.1 Pulse Conditioning — 74C4538W Monostable Multivibrators (U43, U44)

The raw LADC comparator outputs are narrow pulses at 1.52 MHz. In parallel with counting, each pair of channels passes through one dual monostable multivibrator (74C4538W). U43 handles channels 5 and 6 (LADC5/LADC6); U44 handles channels 7 and 8 (LADC7/LADC8). The 74C4538W is a retriggerable, resettable one-shot: on each rising edge from the comparator, it generates a clean output pulse of fixed width, set by an external RC network (C141–C144, 0.1 µF). This standardizes pulse width and eliminates the effect of comparator output skewing and is used for visualization of event rate. A red LED (D22–D25, current-limited by 270 Ω resistors) on each channel output provides a visible activity indicator during detector commissioning and diagnostics.

### 9.2 Microcontroller — Teensy 4.1 (U1)

The central processing element is a Teensy 4.1 development module (U1), based on the NXP i.MX RT1062 ARM Cortex-M7 microcontroller running at 600 MHz The Teensy 4.1 was chosen for its combination of high clock speed, abundant I/O, hardware-accelerated serial peripherals, and compact footprint. The key functions it performs in this system are:

- **Pulse counting.** The Teensy reads LADC5–LADC8 on dedicated digital timer input pins. For each channel, it counts the number of comparator pulses during a single ring-down event. This count is the logarithmic digital word representing the input charge.
- **Threshold control.** Via SPI, the Teensy writes to the TLV5626 4-channel DAC (U47), setting the per-channel threshold voltages THR5–THR8. This allows software-programmable gain calibration on a per-channel basis without any manual adjustment.
- **Data formatting and transmission.** After each event, the Teensy packages the 4-channel count values and stores them. The 1-minute accumulated counts together with 1-minute accumulated spectrum are stored on SD card.
- **System monitoring.** Status LEDs (LED1 red, LED2 yellow/green via U3 area) provide visual feedback event rates.

The Teensy exposes SPI (D10_CS, D11_MOSI, D12_MISO, D13_SCK), I²C (D20_SDA, D21_SCL), and UART (RX0/TX0) peripherals. The SPI bus connects to the DAC; the UART is bridged to RS-485 via U3 for host communication. Bulk decoupling capacitors C163 and C167 (47 µF each) on the +5 V and +3.3 V rails suppress supply transients generated by the microcontroller’s switching activity.

### 9.3 Programmable Threshold DAC — TLV5626DW (U47)

The TLV5626DW is an 8-channel, 8-bit serial DAC with an SPI-compatible interface. It receives digital threshold codes from the Teensy and outputs 4 analog voltages (THR5–THR8) that directly bias the inverting inputs of the four LMV7219 comparators via the THR5–THR8 lines. The reference voltage for the DAC is set by REF1, derived from the +3.3 V supply through a resistor divider (R188, R189), fixing the full-scale DAC output range.

This architecture confers important advantages. First, the threshold for each channel can be tuned individually to compensate for channel-to-channel gain variations in the analog front-end, equalizing the effective detection threshold across all eight inputs. Second, thresholds can be adjusted in real time by software (serial terminal or web interface) without interrupting acquisition, allowing adaptive sensitivity control as cosmic ray flux or background conditions change. Third, threshold sweeping enables in-situ S-curve calibration: by scanning THR while counting events, the system can map the efficiency curve of each channel and extract the optimal operating point.

The LDAC (latch DAC) pin is driven by Teensy to update all four DAC outputs simultaneously, avoiding transient threshold mismatches between channels during an update. Capacitor C181 (0.1 μF) decouples the DAC supply locally.

### 9.4 RS-485 Communication Interface — RS485-HW519 (U3)

Data from the Teensy can be transmitted over an RS-485 differential serial link using the RS485-HW519 transceiver (U3). RS-485 was selected over single-ended alternatives (RS-232, USB) for several reasons characteristic of physics experiment environments:

- **Noise immunity.** RS-485 uses differential signaling with typical common-mode rejection exceeding 25 dB, critical in environments where high-voltage power supplies, motors, and switching regulators generate significant electromagnetic interference.
- **Cable length.** RS-485 supports cable runs up to 1200 m at 100 kbit/s, enabling the detector module to be located at the measurement point while the host computer resides in a shielded rack room.
- **Multi-drop bus.** A single RS-485 bus can address up to 32 nodes, allowing multiple LADC modules to share a single cable to the host, simplifying wiring in multi-detector arrays.

The transceiver connects its TXD and RXD pins directly to the Teensy UART, with the driver enable controlled by a GPIO pin. The A/B differential pair exits to the field cabling. Termination resistors (not shown in schematic) are assumed to be fitted at the cable endpoints.

### 9.5 Ethernet Network Services — HTTP, FTP, and NTP (Teensy 4.1)

The Teensy 4.1 includes a built-in 10/100 Mbit/s Ethernet MAC with an on-board magnetics connector, enabling the LADC module to participate directly in a laboratory LAN without any additional network hardware. Three complementary network services are implemented in firmware.

An embedded HTTP server exposes a lightweight REST-style interface: a host browser or DAQ program can issue simply GET requests to retrieve the latest 4-channel event record in JSON format, query per-channel threshold settings, or push new threshold values via POST, providing a human-readable diagnostic interface with no dedicated client software required.

An FTP server complements this by offering file-based access to accumulated event logs stored on the Teensy's micro-SD card, allowing a remote operator to download multi-hour data runs as plain binary or CSV files using any standard FTP client without interrupting ongoing acquisition.

Accurate event timestamping is provided by an NTP client that periodically queries a laboratory or public NTP server and disciplines the Teensy's real-time clock; because cosmic ray coincidence analysis requires sub-millisecond timestamp agreement between geographically separated detector stations, NTP synchronization is essential for correlating events recorded by different LADC modules within an array. Together, these three services transform the module from a standalone serial instrument into a network-addressable sensor node that can be integrated into a distributed cosmic ray observatory with no specialized infrastructure beyond a standard Ethernet switch.

### 9.5.1 HTTP Server Web Interface

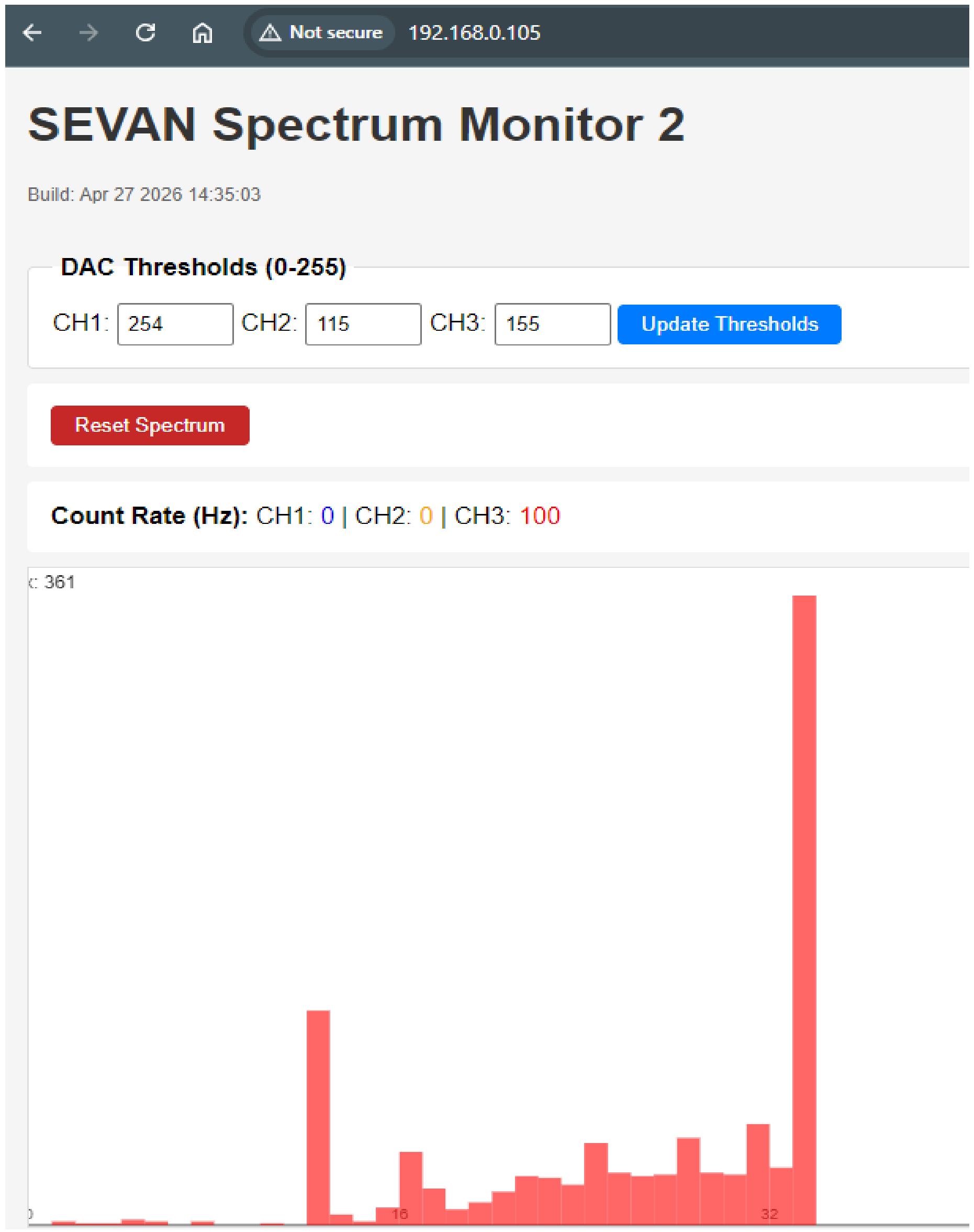

The SEVAN2 firmware on a Teensy 4.1 microcontroller exposes a lightweight HTTP server on port 80 via QNEthernet. It provides a real-time browser dashboard for monitoring 4-channel particle detector data and a JSON API for programmatic access. No additional software is required — any modern web browser can connect directly to the device IP.

#### API Routes

| Route | Method | Description |
|---|---|---|
| / | GET | HTML dashboard. Optional query params: t1–t4 (threshold 0–255), reset=1 (clear spectrum data) |
| /api/data | GET | JSON payload with live frequency counts and 128-bin spectrum arrays for all 4 channels. Polled every 1 second. |

#### Dashboard Features

| Element | Details |
|---|---|
| **Count Rates** | Live pulse frequency in Hz for CH1 (blue), CH2 (orange), CH3 (red), CH4 (green) |
| **DAC Thresholds** | 4 numeric inputs (0–255) submitted as GET params t1–t4 to set discriminator levels per channel |
| **Energy Spectrum** | Canvas bar chart of 128-bin pulse-height distributions, all 4 channels overlaid and color-coded |
| **Status Indicator** | Green (online) / Red (offline) badge based on API reachability; firmware build timestamp displayed |
| **Reset Button** | Clears all spectrum histograms and counters via GET /?reset=1 |

#### JSON Data Format (/api/data)

```
{  "freq":  [Hz_CH1,  Hz_CH2,  Hz_CH3,  Hz_CH4],    "data0":  [...128  bins...],
"data1": [...],  "data2": [...],  "data3": [...] }
```

Each dataN array contains 128 integers representing the pulse-height histogram for channel N+1. The freq array holds the current count rate in Hz for each channel.

#### Network Configuration

Network settings are read from **/settings.txt** on the SD card at boot. Supported keys: mac, ip, mask, gateway, dns, ntp, timezone. Missing keys fall back to compiled-in defaults.

An FTP server is also available on port 21 (credentials: user crd / password crd) for direct SD card file access and settings management.

### 9.6 Power Supply

The module requires four external supply rails delivered via connector J16: ±15 V (analogue) and +5 V. Two linear regulators convert the ±15 V rails to the ±12 V required by the LM6171BIM op-amps and the output stage transistors:

| Regulator | Conversion | Function |
|---|---|---|
| **US1 (LM7812CT)** | +15 V → +12 V | Positive analogue supply for op-amp stages |
| **UA8 (LM7912CT)** | −15 V → −12 V | Negative analogue supply for op-amp stages |

Input capacitors C182/C183 (0.1 µF) and C180 (33 µF) filter incoming supply noise before the regulators; output capacitors C164/C179 (1 µF) ensure regulator stability and suppress high-frequency transients. The +5 V and +3.3 V rails are supplied externally (from a system-level regulator board) and are locally decoupled by C163 and C167 (47 µF each). The use of linear regulators for the ±12 V rails is deliberate: switching regulators would inject noise at their switching frequency into the sensitive analog supply, degrading the signal-to-noise ratio of the LC ring-down measurement.

### 9.7 Complete Signal and Data Flow

The end-to-end operation of the LADC module for a single cosmic ray event is as follows. A particle traverses a scintillator or other charge-producing detector element connected to one of the four PIN inputs. The resulting charge pulse excites the corresponding LC tank into free oscillation. The ring-down waveform is amplified by the three-stage op-amp chain and presented to the LMV7219 comparator, which generates a burst of digital pulses. The burst duration — and hence pulse count — is proportional to the logarithm of the deposited charge. The Teensy 4.1 MCU counts the pulses on each active channel. At the end of the ring-down (detected as a cessation of pulses for a set timeout period), the Teensy formats the event record containing the channel counts and stores the result. The DAC-set thresholds remain static between events unless the host issues a threshold update command.

#### 9.7.1 PCB

The PCB of LADS is 2-layer 110x160x1.5mm FR4. Inputs are standard BNC connectors.

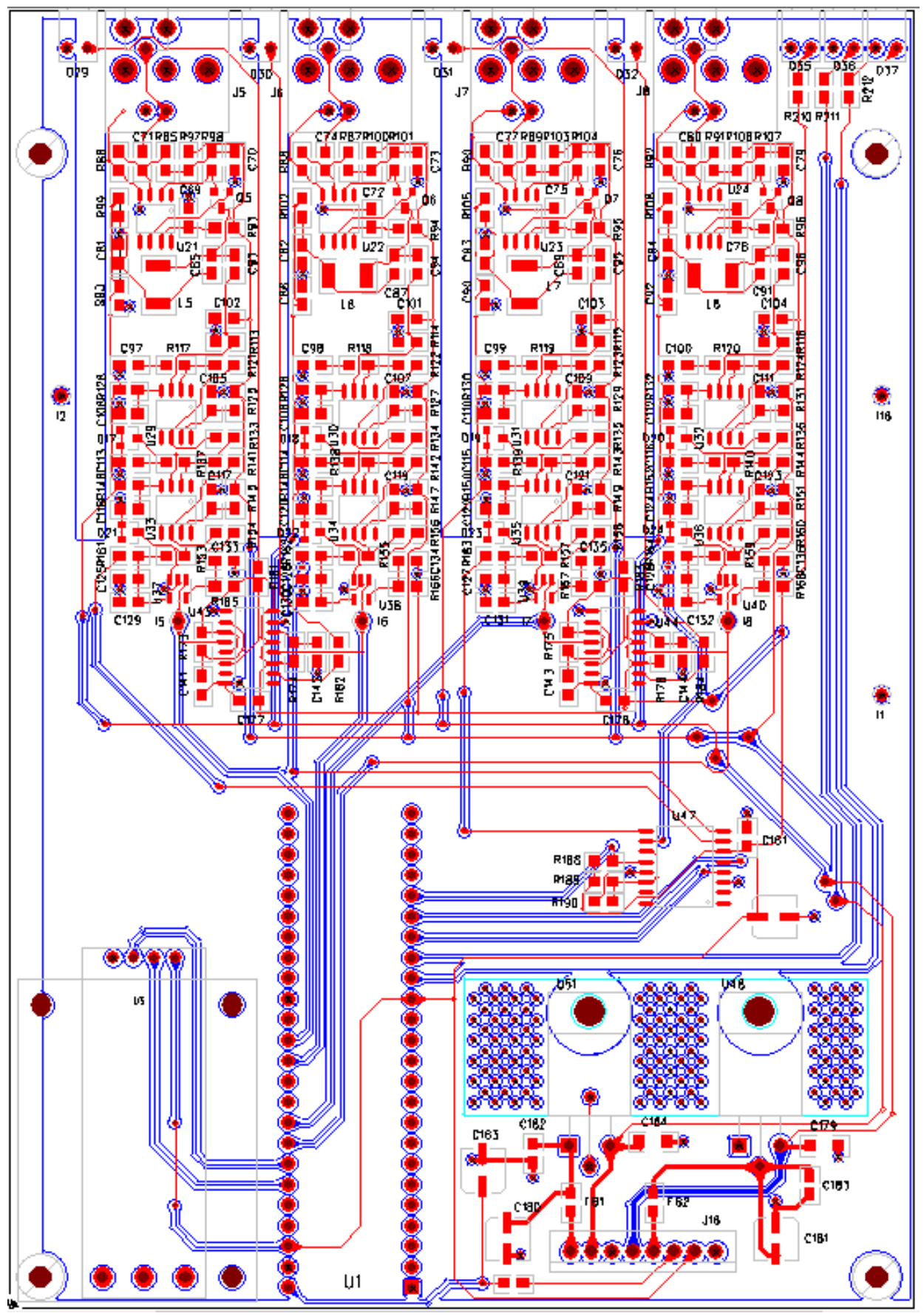

## 10. Physics Motivation and Scientific Applications of SEVAN Light Spectrometer

The SEVAN Light spectrometric detector was developed to provide simultaneous measurements of charged and neutral secondary cosmic-ray components in two key domains of modern high-energy atmospheric and solar physics: Ground Level Enhancements (GLEs, Greider, 2001; Gopalswami, 2012) produced by relativistic solar protons, and Thunderstorm Ground Enhancements (TGEs, Chilingarian et al., 2010; 2011)) produced by Relativistic Runaway Electron Avalanches (RREA) within thundercloud electric fields. Unlike conventional neutron monitors or standard scintillation detectors that mainly provide integral count rates, SEVAN Light combines:

- energy-deposit spectroscopy,
- coincidence topology,
- particle-type separation,
- and high dynamic range electronics.

A major advantage of SEVAN Light is its broad recoverable energy range. The logarithmic ADC electronics enables reconstruction of deposited energies from approximately 5 MeV to 300 MeV, substantially exceeding the dynamic range of standard linear spectrometric systems. This wide range covers:

- relativistic electrons and gamma rays generated in RREA avalanches,
- atmospheric neutrons,
- minimum-ionizing muons,
- and high-energy secondary bundles from solar proton cascades.

SEVAN Light (Space Environment Viewing and Analysis Network, Chilingarian et al., 2009; 2024a) is a compact, energy-deposit-resolving particle spectrometer that can separate neutral components (neutrons, γ-rays) from charged components (muons, electrons) of GLEs and TGEs (Sargsyan and Chilingarian, 2026a; Chilingarian et al., 2026b). SEVAN Light detectors installed at Aragats (3200 m) include a 0.25 m² spectrometric scintillator with a 20 cm thickness and a 1 cm thick, 1 m² veto scintillator (see Fig. 1), with an energy threshold of ≈ 5 MeV. The detector incorporates high-dynamic-range logarithmic ADCs (LADCs) to measure deposits of 5–300 MeV and coincidence logic to classify events as charged or neutral particles. The vertical coincidence window is 1 μs, effectively separating accidental coincidences. SEVAN detectors are among the few instruments capable of simultaneously measuring count rates and energy spectra of both charged and neutral cosmic ray species. The list of available information from SEVAN Light is as follows:

- 1-minute count rates of stacked 1 and 20-cm thick scintillators.

- 1-minute count rates of the coincidences “01”, signal only in the 20 cm scintillator; “10” – signal only in the upper 1-cm thick scintillator, and “11” – signal in both scintillators.

- 1-minute histograms of energy deposits in both scintillators. Histograms corresponding to the coincidences mentioned above are stored continuously over the years of operation.

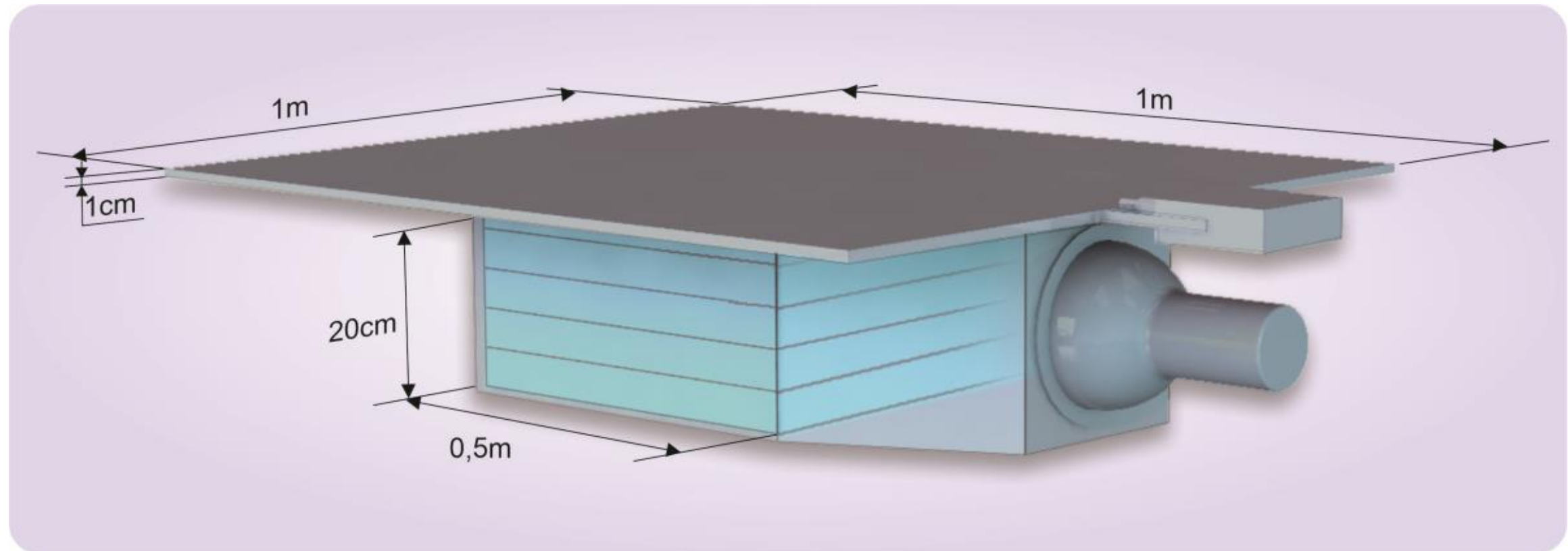

Figure 1. Setup of SEVAN Light detector. 20 cm thick 0.25 $m^2$ area spectrometric scintillator, and 1cm thick, 1 $m^2$ area “veto” scintillator

The samples selected by the "01" coincidence, which generate a signal in the lower scintillator but not in the upper, will be enriched in neutral particles. The samples selected by the “11” coincidence will be enriched in charged particles. In addition, the "10" coincidence will select mostly low-energy charged particles that stop before reaching the lower scintillator. The “11” coincidence channel is predominantly governed by TGE electrons, with a minor contribution from gamma rays.

This simple but highly efficient classification scheme is extremely important because it allows two fundamentally different physical problems to be studied with the same detector.

For solar physics and space-weather applications, SEVAN Light separates the neutron- and muon-enriched components of Ground Level Enhancements (Chilingarian et al., 2024a; 2024c). During GLEs, high-energy solar protons interact with the atmosphere, producing secondary cascades that include neutrons, muons, electrons, and gamma rays. The relative abundance and spectral shape of neutrons and muons depend strongly on the hardness of the primary solar proton spectrum. Consequently, simultaneous neutron–muon spectrometry provides a direct diagnostic of the highest-energy component of solar energetic particles.

This capability is especially important for operational space weather, as the most hazardous solar protons are those with energies above approximately 50–100 MeV, which can damage spacecraft electronics, aviation systems, electrical grids, and pipelines. Traditional neutron-monitor networks provide only integral counting rates and are limited by strong geomagnetic cutoffs. In contrast, SEVAN Light measures energy-dependent observables locally and directly, allowing estimation of spectral hardness in near real time. The detector, therefore, acts as a compact spectrometric analyzer of solar proton cascades. In Figure 3, we present the recovered differential energy spectra of muons (coincidence “11”) and neutrons (coincidence “01”) of the first and second peaks of spectra registered during the largest in 25 years ground level enhancement GLE 77.

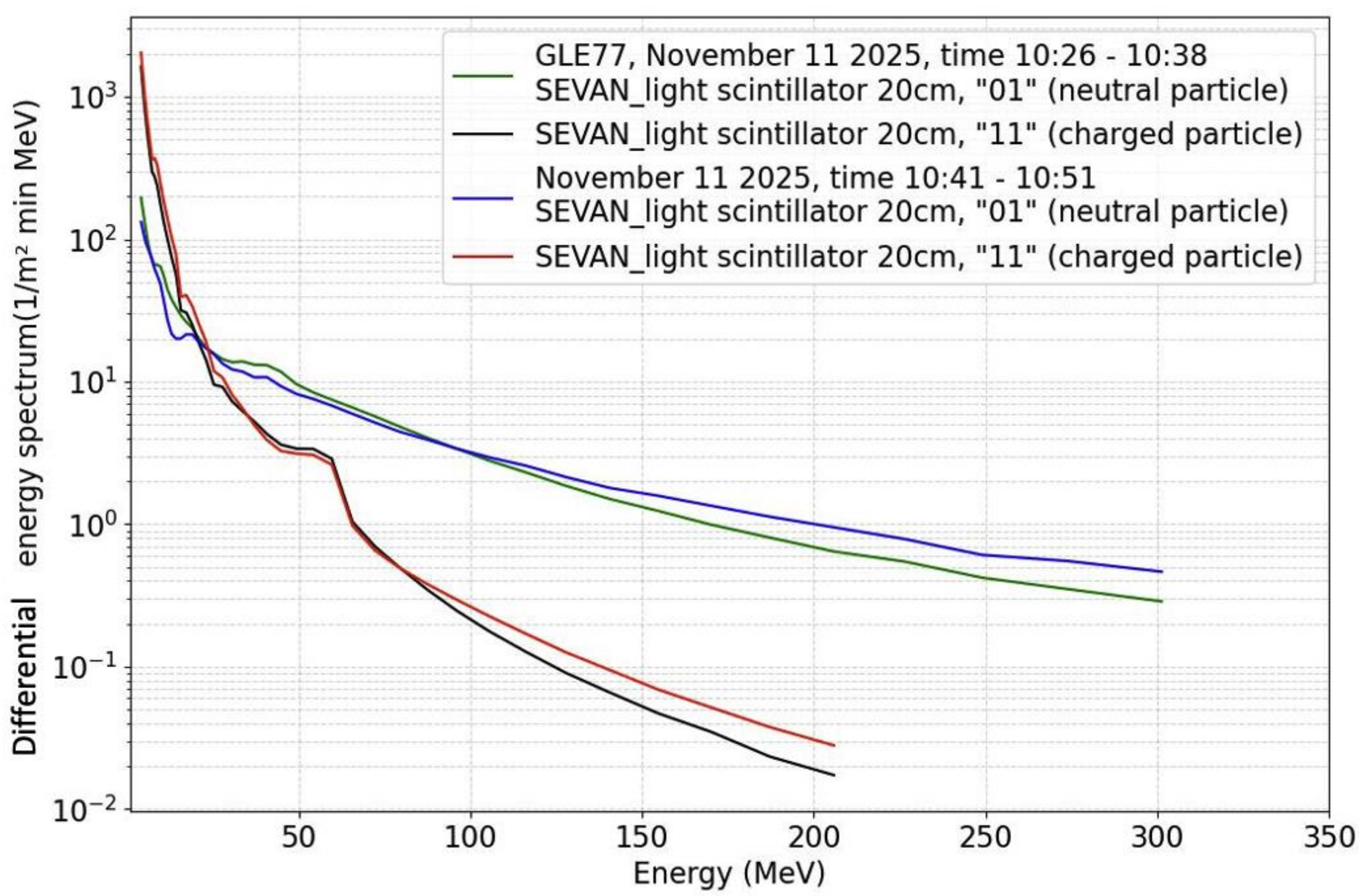

Figure 2. Recovered differential spectra of muon- and neutron-enriched components for the two GLE 77 peaks.

In atmospheric high-energy physics, SEVAN Light provides equally important capabilities for studying TGEs and RREA avalanches. In thunderstorm electric fields, relativistic electrons accelerate and produce bremsstrahlung gamma rays. The physical interpretation of TGEs critically depends on the reliable separation of electron and gamma ray fluxes (Chilingarian et al 2024c; 2024d). SEVAN Light performs this task naturally through coincidence topology and energy-deposit spectra. The “11” coincidence selects penetrating electrons, while the “01” coincidence enriches neutral gamma-ray samples. This enables simultaneous reconstruction of electron and gamma-ray energy spectra during thunderstorm particle bursts. Figure 2 displays 2 of the largest recent TGEs with high electron content, showing the “11” and “01” coincidences of SEVAN Light, lightning distance, temperature, and dew point, and near-surface electric field. The figure indicates that electron and gamma enhancements occur during periods of intense electric-field activity, and are often abruptly ended by a lightning flash

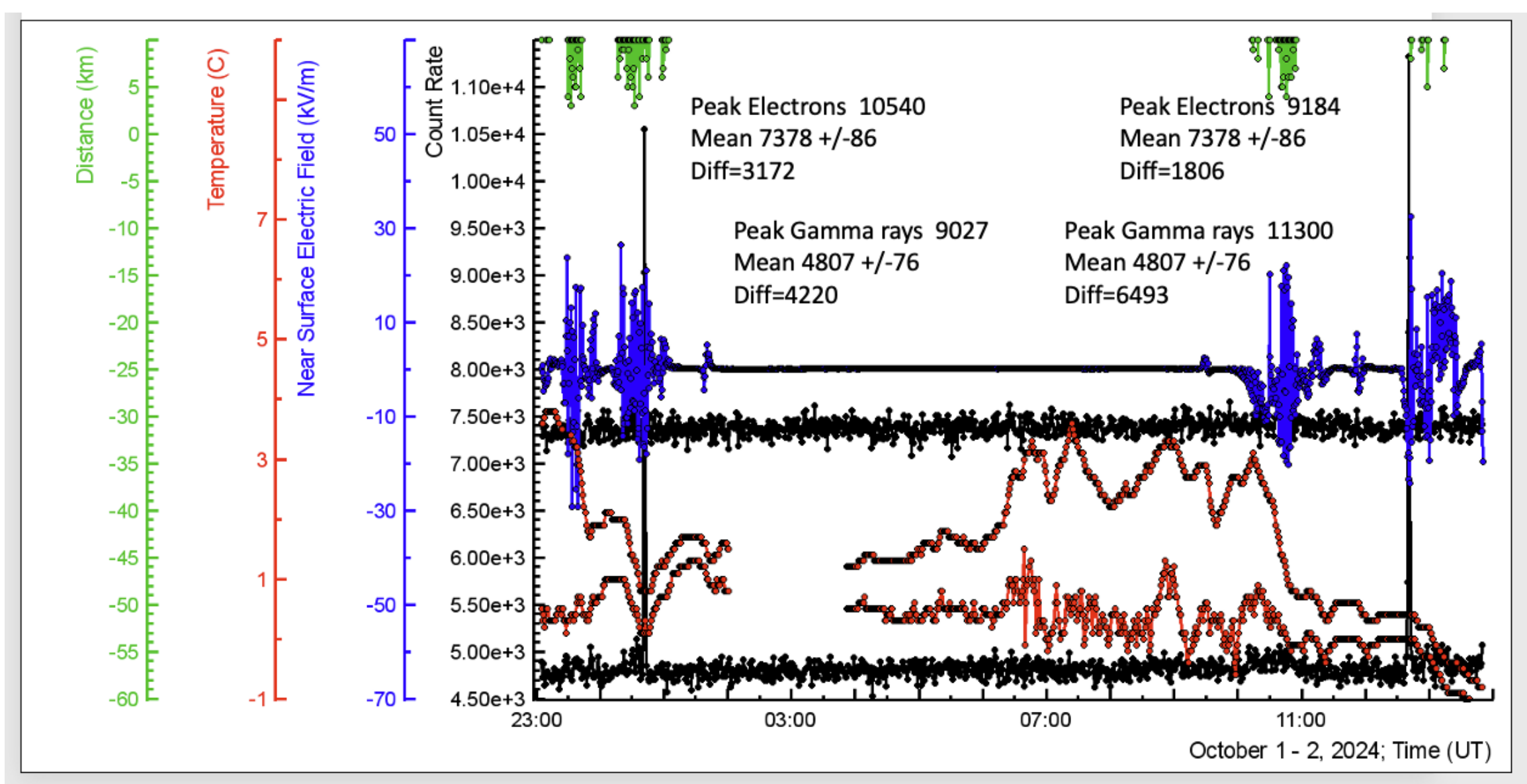

Figure 3. Electron-rich TGEs: each event is shown with natural count rate, lightning distance, temperature, dew point, and NSEF. The selected events occur in a specific meteorological and electrical environment, usually with low cloud base and nearby lightning.

2 October 2024 at 00:42 UT is a very strong gamma event. The gamma enhancement reaches 118.8% with 64.8 σ significance, while the electron channel increases by 22.3% with 19.0 σ significance. The cloud base is at 87.5 m, but it is still extremely low compared with ordinary TGEs.

## Conclusion

The LADC module, designed at the Yerevan Physics Institute for cosmic ray detection, demonstrates a complete and elegant approach to wide-dynamic-range charge measurement. Four parallel logarithmic ADC channels — each based on LC tank ring-down counting — deliver inherently logarithmic charge-to-digital conversion without dedicated log amplifiers or complex nonlinear circuits. The digital back end, built around a Teensy 4.1 ARM Cortex-M7 microcontroller, provides pulse counting, per-channel threshold control via a TLV5626 DAC, and RS-485 serial output for long-distance noise-immune communication to a host data acquisition system. Via Ethernet interface it hosts an HTTP server for interface and control and FTP server for stored data management. Linear voltage regulators preserve analogue supply purity. The result is a self-contained, calibratable, multi-channel detector readout module suited to the harsh electromagnetic environment of cosmic ray observatories, where signal amplitudes span several decades and cable distances can reach hundreds of meters.

An especially important advantage is that the detector can measure both weak and extremely intense TGEs without saturation. Large TGEs on Aragats can increase particle fluxes by several factors above the cosmic-ray background. Conventional systems often lose spectrometric information during such bursts, whereas the logarithmic ADC preserves spectral reconstruction over a very broad dynamic range. This makes SEVAN Light uniquely suited to studying the evolution of RREA spectra under rapidly changing atmospheric electric fields. Because the veto scintillator area is larger than the spectrometric scintillator area and the efficiency of gamma-ray conversion in a 1-cm-thick scintillator is low, the charged-particle veto efficiency approaches 100%. This strongly suppresses contamination by gamma rays and inclined muons that can mimic neutral particles in thick detectors. Thus, SEVAN Light combines several critical physical advantages:

1. Broad recoverable energy range (≈5–300 MeV),
2. Simultaneous charged/neutral particle separation,
3. Recovery of differential energy spectra,
4. High-rate capability without saturation,
5. Reliable neutron–muon classification for solar physics,
6. Reliable gamma–electron classification for atmospheric high-energy physics,
7. Compact geometry is suitable for distributed detector networks.

These capabilities make SEVAN Light a versatile instrument for studying both cosmic and atmospheric particle accelerators. In solar physics, it provides rapid diagnostics of relativistic solar proton spectra relevant to space-weather alerting. In atmospheric physics, it enables detailed investigation of relativistic runaway electron avalanches and high-energy radiation from thunderstorms. The same detector architecture, therefore, links two major natural particle acceleration environments: the Sun and Earth's atmosphere.

## Conflict of Interest

The authors declare no conflicts of interest relevant to this study.

## Data Availability Statement

The TGE archive through 2024 is freely available in graphical and numerical formats from the Mendeley site at DOI: 10.17632/8gtdbch59z (Chilingarian et al., 2024c). The online data on continuous monitoring of meteorological parameters, electric field measurements, and particle fluxes at Aragats are available in graphical and numerical formats on the ADEI platform of CRD YerPhI: http://crd.yerphi.am/adei/